 \documentclass[twocolumn,prl,aps,psfig,showpacs,preprintnumbers,superscriptaddress]{revtex4}
\usepackage{graphicx}
\usepackage{epstopdf}
\usepackage{float}
\usepackage{color}
\usepackage{amsmath}

\begin{document}

\title{Ferromagnetic spin fluctuations in the filled skutterudite SrFe$_4$As$_{12}$ revealed by $^{75}$As NMR-NQR measurements}
\author{Q.-P. Ding}
\affiliation{Ames Laboratory, U.S. DOE, and Department of Physics and Astronomy, Iowa State University, Ames, Iowa 50011, USA}
\author{K. Rana}
\affiliation{Ames Laboratory, U.S. DOE, and Department of Physics and Astronomy, Iowa State University, Ames, Iowa 50011, USA}
\author{K. Nishine}
\affiliation{Muroran Institute of Technology, Muroran, Hokkaido 050-8585, Japan}
\author{Y. Kawamura}
\affiliation{Muroran Institute of Technology, Muroran, Hokkaido 050-8585, Japan}
\author{J. Hayashi}
\affiliation{Muroran Institute of Technology, Muroran, Hokkaido 050-8585, Japan}
\author{C. Sekine}
\affiliation{Muroran Institute of Technology, Muroran, Hokkaido 050-8585, Japan}
\author{Y. Furukawa}
\affiliation{Ames Laboratory, U.S. DOE, and Department of Physics and Astronomy, Iowa State University, Ames, Iowa 50011, USA}

\date{\today}

\begin{abstract} 

       $^{75}$As nuclear magnetic resonance (NMR) and nuclear quadrupole resonance (NQR) measurements have been carried out to investigate the magnetic and electronic properties of the filled skutterudite metallic compound SrFe$_4$As$_{12}$.
   The temperature dependence of Knight shift $K$ determined by the NQR spectrum under  a small magnetic field ($\le$ 0.5 T) shows the similar temperature dependence of the magnetic susceptibility $\chi$ which exhibits a broad maximum at $T^\ast$ $\sim$ 50 K. 
    The nuclear spin-lattice relaxation rate divided by temperature, 1/$T_1T$, increases with decreasing temperature and exhibits a broad maximum at $T$ $\sim$ 70 K,  similar to the case of $\chi$. 
    The temperature dependence of $K$ and $1/T_1T$ is reasonably explained by a simple model where we assume a concave-shaped band structure near the Fermi energy.
     Based on a Korringa ratio analysis using the $T_1$ and $K$  data,  ferromagnetic spin fluctuations are found to exist in SrFe$_4$As$_{12}$.  
    These results indicate that SrFe$_4$As$_{12}$ can be characterized to be a metal with ferromagnetic correlations and also  the peculiar band structure responsible for the suppression of $1/T_1T$ and $K$  at low temperatures.

\end{abstract}

%  \pacs{74.70.Xa, 76.60.-k}
\maketitle

 \section{I.  Introduction} 
 
 %    Magnetic fluctuations are one of the key parameters to characterize strongly correlated materials.
%   Because these compounds are difficult to prepare, many physical properties of these mate rials are still under investigation. 
     The investigation of physical properties of filled skutterudite-structure compounds has a long history which started with the discovery of  LaFe$_4$P$_{12}$ by Jeitschiko {\it et~al.}  in 1977 \cite{Jeitschko1977}.
     The compound crystallizes in a body-centered cubic structure with the space group $Im\overline{3}$, and is now known to have a wide variety of compounds with the general formula $AT_4X_{12}$ ($A$ = alkali metal, alkaline earth metal, lanthanide, and actinide; $T$ = Fe, Ru, Os, Co, Rh, Ir, and Pt; $X$ = P, As, Sb, and Ge).
    These compounds have received much attention since  they exhibit rich physical properties that include  superconductivity, metal-insulator transition, ferromagnetism, antiferromagnetism, hybridization gaps (Kondo insulator behavior), nonfermi-liquid behavior, quadrupolar ordering, and field-induced heavy-fermion states \cite{Sales2005,Shirotani1997,Sekine1998,Leithe-Jasper2003,Schnelle2008,Ishida2005,Matsuoka2005,Matsumura2005,Nakai2005,Matsumura20071,Shimizu2007,Tou2011,Magishi2014,Kawamura2018}, as well as excellent thermoelectric properties \cite{Sales1996}.

    These intriguing physical phenomena are mainly owing to  $f$ electrons in rare-earth or $d$ electrons in transition metals hybridized with $p$ electrons of $X$ elements, as well as the characteristic cage structure in the compounds.
    Many studies in rare-earth-filled skutterudite compounds have been carried out to investigate the role of the $f$-electrons on physical properties.
    Instead, few studies on $d$ electron systems have been performed so far.
    Most studies on $d$ electron systems focus on  iron-antimony filled skutteudite compounds $A$Fe$_4$Sb$_{12}$ and found  the interesting magnetic properties which largely depend on the number of valence electrons of the $A$ ions.
      In the case of $A$ = monovalent Na and K ions, a weak ferromagnetism with a Cuire temperature of $T_{\rm C}$ = 85 K has been observed \cite{Leithe-Jasper2003,Leithe-Jasper2004}, while no magnetic order has been reported for the case of divalent alkaline-earth ions such as Ca, Sr, and Ba, where ferromagnetic spin fluctuations are considered to  play an important role \cite{Schnelle2005,Schnelle2008,K}.
      On the other hand, the importance of antiferromagnetic spin fluctuations are pointed out in the trivalent ion system of LaFe$_4$Sb$_{12}$  \cite{Schnelle2008,Gippius2006}.
     Although it would be  important to systematically study these physical properties of $d$ electron systems by changing $X$ ions such as P and As for further deep understandings of the role of $d$ electrons, not much studies have been carried out  because of the difficulty in preparing those compounds.

     Recently, new filled-skutterudite arsenide compounds  Sr$T_4$As$_{12}$ ($T$ = Fe, Ru, Os) have been synthesized using a high-pressure synthesis technique \cite {Nishine2017}, which provides a new opportunity of systematic studies of the role of $d$ electrons with different $X$ ions as well as the effects from the different $d$ electron of  3$d$, 4$d$ and 5$d$.
     For 5$d$ electron system,  SrOs$_4$As$_{12}$ was  found to  be a new superconductor with a transition temperature of $T_{\rm c}$ = 4.8 K  \cite {Nishine2017}.
     For 3$d$ and 4$d$ electron systems, on the other hand, SrFe$_4$As$_{12}$  and  SrRu$_4$As$_{12}$ do not exhibit superconductivity down to 2 K, although the electrical resistivities  show metallic behavior \cite{Nishine2017}. 
     As for the magnetic properties for 3$d$ systems, comparative studies have been performed. 
      The static magnetic susceptibility $\chi$ of the isostructural compound BaFe$_4$As$_{12}$ shows a broad maximum at $T^\ast$ $\sim$ 50  K.  
      Above $T^\ast$, $\chi$ follows the Curie-Weiss (CW) law with an effective magnetic moment $\mu_{\rm eff}$ of 1.46 $\mu_{\rm B}$/Fe and a negative Weiss temperature $\theta$ of -57 K \cite{Sekine2015}.
% suggesting ferromagnetic spin correlations. 
%    A similar behavior of $\chi$ has been reported in the weak itinerant ferromagnet LaFe$_4$Sb$_{12}$ \cite{Matsumura20071}. 
    Together with the observation of  a large Sommerfeld coefficient of the electronic specific heat ($\gamma$ = 62 mJ mol$^{-1}$K$^{-2}$),  Sekine ${\it et~al.}$  suggested that   BaFe$_4$As$_{12}$ is a nearly ferromagnetic metal \cite{Sekine2015}.
     On the other hand,  in the case of SrFe$_4$As$_{12}$, although the static magnetic susceptibility exhibits a broad maximu at $T^\ast$ $\sim$ 50 K and follows the CW law with $\mu_{\rm eff}$ = 1.36 $\mu_{\rm B}$/Fe very similar to the case of BaFe$_4$As$_{12}$,  the positive Weiss temperature $\theta$ of 36 K has been reported,  indicating a different magnetic spin correlations in comparison with the case of BaFe$_4$As$_{12}$.
      A large $\gamma$ = 58 mJ mol$^{-1}$K$^{-2}$ is also observed in  SrFe$_4$As$_{12}$  which suggests strong electron correlation effects, as well as the case of  BaFe$_4$As$_{12}$. 
 %     Thus it is important and interesting to investigate the electronic and magnetic properties of SrFe$_4$As$_{12}$. 

    Motivated by the reported interesting magnetic properties in SrFe$_4$As$_{12}$, we have carried out nuclear magnetic resonance (NMR) and nuclear quadrupole resonance (NQR) measurements which are powerful techniques to investigate the magnetic and electronic properties of materials from a microscopic point of view.
    It is known that the temperature dependence of the nuclear spin-lattice relaxation rate (1/$T_1$) reflects the wave vector $q$-summed dynamical susceptibility. 
   On the other hand, NMR spectrum measurements, in particular the Knight shift $K$, give us information on static magnetic susceptibility $\chi$. 
   Thus from the temperature dependence of 1/$T_1T$ and $K$, one can obtain valuable insights about magnetic fluctuations in materials. 
      In this paper, we report the results of $^{75}$As NMR and NQR measurements performed for investigating the spin fluctuations in SrFe$_4$As$_{12}$.
    Our analysis, based on the modified Korringa relation, reveals  electron correlations enhanced  around ferromagnetic (FM) wavenumber $q$ = 0  in SrFe$_4$As$_{12}$.
   The characteristic  temperature dependence of  $K$  and 1/$T_1T$ was reasonably explained by a simple model where  a concave-shaped band structure near the Fermi energy is assumed.

 %  The positive $\theta$ in the CW susceptibility above 150 K is favor to a ferromagnetic ordering at finite temperature.
%    However no definitive evidence for a long range order has been found.

 \section{II.  Experimental}
 
    Polycrystalline SrFe$_4$As$_{12}$ samples  were prepared at high temperatures and high pressures using a Kawai-type double-stage multianvil high-pressure apparatus \cite {Nishine2017}. 
    NMR and NQR measurements of $^{75}$As ($I$ = $\frac{3}{2}$, $\frac{\gamma_{\rm N}}{2\pi}$ = 7.2919 MHz/T, $Q=$ 0.29 barns) nuclei were conducted using a lab-built phase-coherent spin-echo pulse spectrometer.
   The $^{75}$As-NMR spectra were obtained by sweeping the magnetic field $H$ at a fixed frequency $f$ = 37 MHz, while $^{75}$As-NQR spectra were measured in steps of frequency by measuring the intensity of the Hahn spin echo.  
   The $^{75}$As nuclear spin-lattice relaxation rate (1/$T_{\rm 1}$) was measured with a saturation recovery method.
   $1/T_1$ at each temperature ($T$) was determined by fitting the nuclear magnetization $M$ versus time $t$  using the exponential function $1-M(t)/M(\infty) = e^ {-3t/T_{1}}$ for $^{75}$As NQR,  where $M(t)$ and $M(\infty)$ are the nuclear magnetization at time $t$ after the saturation and the equilibrium nuclear magnetization at $t$ $\rightarrow$ $\infty$, respectively. 

  \section{III.  Results and discussion}
 \subsection{A.  $^{75}$As NMR and NQR spectra}

\begin{figure}[tb]
\includegraphics[width=\columnwidth]{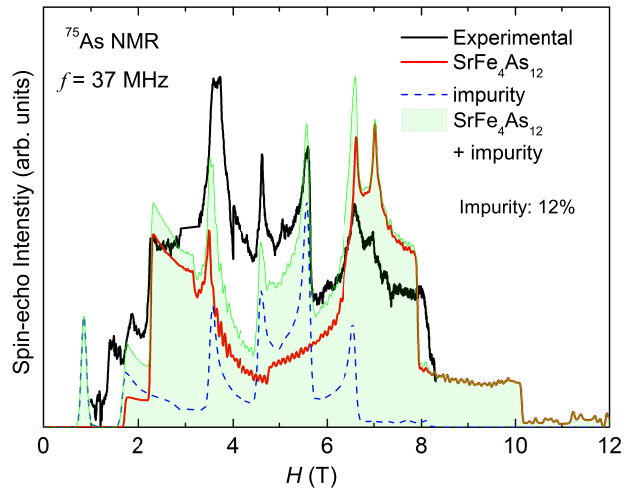} 
\caption{Field-swept $^{75}$As-NMR spectra of SrFe$_4$As$_{12}$ at $f$ = 37 MHz and $T$ = 4.3 K.
Black curve is the observed spectrum and red curve is the calculated spectrum with $\nu_{\rm Q}$ = 54.8 MHz, $\eta$ = 0.4. %  and $K$ = 0.5 $\%$. 
Blue dotted curve represents the calculated $^{75}$As NMR spectrum ($\nu_{\rm Q}$ = 23.5 MHz, $\eta$ = 0) from the impurity phase \cite{impurity}.
The sum of the two calculated spectra is shown by the green area. }
\label{fig:NMR}
\end{figure}   

      Figure\ \ref{fig:NMR} shows the field-swept $^{75}$As-NMR spectrum in SrFe$_4$As$_{12}$ at $T$ = 4.3 K. 
     The typical NMR spectrum can be described by a nuclear spin Hamiltonian which is  a sum of the nuclear Zeeman  (${\cal H}_{\rm M}$)  and electric quadrupole (${\cal H}_{\rm Q}$) interactions:
%  \begin{align}
% {\cal H}& ={\cal H}_{\rm Z}+{\cal H}_{\rm Q}\nonumber \\
% &=-\gamma\hbar(1+K)H[\frac{1}{2}(I_+ e^{-i\phi}+I_- e^{i\phi}){\rm sin}\theta+I_Z{\rm cos}\theta]\nonumber \\
% & +\frac{h \nu_{\rm Q}}{6} [3I_{Z}^{2}-I^2 + \frac{1}{2}\eta(I_+^2 +I_-^2)] ,
% \label{eq:hamitonian}
% \end{align}
\begin{equation}
 {\cal H} ={\cal H}_{\rm M}+{\cal H}_{\rm Q}, 
\label{eq:hamitonian1}
\end{equation}
where 
\begin{equation}
 {\cal H}_{\rm M} =-\gamma\hbar(1+K)H[\frac{1}{2}(I_+ e^{-i\phi}+I_- e^{i\phi}){\rm sin}\theta+I_Z{\rm cos}\theta] 
\label{eq:hamitonian2}
\end{equation}
and 
\begin{equation}
{\cal H}_{\rm Q} = \frac{h \nu_{\rm Q}}{6} [3I_{Z}^{2}-I^2 + \frac{1}{2}\eta(I_+^2 +I_-^2)] ,
\label{eq:hamitonian3}
\end{equation}
 in the coordinate of the principal $X$, $Y$, and $Z$ axes of electric field gradient (EFG).
Here  $H$ is the applied field, $h$ is Planck's constant, $\nu_{\rm Q}$ is nuclear quadrupole frequency defined by $\nu_{\rm{Q}}$ =  $eQV_{ZZ}$/2$h$ where $Q$ is the quadrupole moment of the As nucleus, $V_{ZZ}$ is the EFG at the As site,  $\eta$ is the asymmetry parameter of the EFG, and $\theta$ and $\phi$ are the polar and azimuthal angles between the direction of the applied field and the $Z$ axis of EFG, respectively. 

      Different from the case of a nuclear spin $I$ = 3/2 with larger Zeeman interaction and small perturbed quadrupolar interaction where a  central  transition and two satellite peaks can be expected, a complex NMR spectrum is observed in SrFe$_4$As$_{12}$.
     This is due to a strong quadrupole interaction  and a finite asymmetric parameter $\eta$ of the EFG tensor at the As site.
     In order to reproduce  the complex NMR spectrum, we calculated a powder-pattern NMR spectrum by diagonalizing exactly the nuclear spin Hamiltonian without using perturbation theory. 
    The calculated spectrum with the NMR frequency $f$ = 37 MHz, NQR frequency $\nu_{\rm Q}$ = 54.8 MHz and  $\eta$ = 0.4 reasonably reproduces  the characteristic shape of  the observed spectrum, as shown by the red curve in Fig.\ \ref{fig:NMR}.
    However, we notice that, in addition to the calculated spectrum (red curve), there is another contribution ($\sim$12 $\%$ spectral weight)  of  $^{75}$As NMR spectrum with $\nu_{\rm Q}$ = 23.6 MHz and $\eta$ = 0 to the total NMR spectrum. 
     This contribution is assigned to the impurity phase of  arsenic metal  \cite{impurity}.

  The principal axis of EFG at the As site cannot be determined from NMR spectrum measurements on the powder compound. 
   Tou {\it et ~al.} \cite{Tou2005} have determined the principal axis of the EFG at the Sb sites in the isostructural PrOs$_4$Sb$_{12}$ compound from NMR measurements using a single crystal, which reports that, although there is one crystallographically equivalent Sb site in the filled skutterudite structure, there are  three different Sb sites with the principal axis  parallel to [100], [010], and [001] of the crystal, respectively, due to the local symmetry of the 24g site of the Sb ions. 
   The same conclusion for the direction of the EFG at the Sb sites in CeOs$_4$Sb$_{12}$ has been reported from Sb NMR using an oriented powder sample \cite{Yogi2009}.  Since the crystal structure of the Sb compounds is the same with that of SrFe$_4$As$_{12}$, we consider the directions of EFG at the As sites are the same.

\begin{figure}[tb]
\includegraphics[width=\columnwidth]{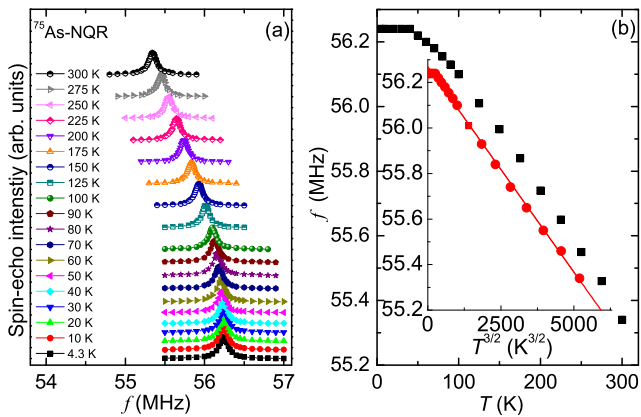} 
\caption{(a) $T$ dependence of the $^{75}$As-NQR spectra for SrFe$_4$As$_{12}$. 
(b) $T$ dependence of $^{75}$As-NQR frequency $\nu_{\rm NQR}$ in SrFe$_4$As$_{12}$.
The inset shows the $^{75}$As-NQR frequency $\nu_{\rm NQR}$ vs. $T ^{3/2}$ plot, indicating a monotonous decrease following $T ^{3/2}$ above 40 K.
}
\label{fig:NQR}
\end{figure}

      In NQR spectrum under zero magnetic field for $I$ = 3/2 (only ${\cal H}_{\rm Q}$ in ${\cal H}$), one expects a single transition line at a frequency of  $\nu_{\rm NQR} = \nu_{\rm Q}\sqrt{1+{\eta}^2/3}$.   
      Using the $\nu_{\rm Q}$ = 54.8 MHz and  $\eta$ = 0.4 at $T$ = 4.3 K estimated from the analysis of the NMR spectrum, one expects the NQR line at $f$ $\sim$ 56.3 MHz which is actually observed as shown in  Fig. \ \ref{fig:NQR} (a). 
    The temperature dependence of the NQR spectra from 4.3 K to 300 K is also presented in the figure.  
       Although the peak position slightly shifts to lower frequency  with increasing temperature above $\sim$ 40 K, the linewidth of the NQR spectra (full-width at half maximum, $FWHM$ $\sim$ 140 kHz)  is nearly independent of temperature and no broadening or splitting has been observed.
 This  indicates that there are no any structural or magnetic phase transitions from 4 K to 300 K in SrFe$_4$As$_{12}$. 
      The temperature dependence of $\nu_{\rm NQR}$ determined from the peak positions of the NQR spectra  is shown in Fig.\ \ref{fig:NQR} (b). 
%      $\nu_{\rm NQR}$ shows almost constant value of 56.24 MHz at low temperatures up to 40 K, then slightly decreases as temperature increases, and decrease more rapidly at higher temperatures.
       Similar temperature dependence of $\nu_{\rm NQR}$ is observed in many filled skutterudite compounds \cite{Matsumura2007,Shimizu2007,Magishi2014,Nowak2009,Nowak2011,Yogi2014} where the temperature dependence at higher temperatures is found to obey an empirical relation $\nu_{\rm NQR}(T) = \nu_{\rm NQR}(0)(1-\alpha_{\rm Q} T^{3/2})$ with a fitting parameter $\alpha_{\rm Q}$.
   This temperature dependence is considered to be due to thermal lattice expansion \cite{alphaQ}.
        As shown in the inset of  Fig.\ \ref{fig:NQR} (b), the temperature dependence of  $\nu_{\rm NQR}$ in SrFe$_4$As$_{12}$ also follows the relation with $\alpha_{\rm Q} = 3.21 \times 10^{-6}$  K$^{-3/2}$.
      The origin of the deviation from the relation below $\sim$ 40 K  is not clear at present, but one of the possible origins could be due to the quench of the isotropic lattice expansion, as has been discussed in Ref. \onlinecite{Matsumura2007}.

 \subsection{B.  $^{75}$As Knight shift}

    Determination of the Knight shift $K$  from the complex NMR spectrum shown in Fig.\ \ref{fig:NMR} is relatively difficult due to the strong quadrupole interaction and relatively large asymmetric parameter $\eta$ value, along with the presence of the impurity phase.
     Usually $K$ is estimated from NMR spectrum which is performed under magnetic field large enough to the quadrupole interaction.
     In the case of SrFe$_4$As$_{12}$, $\nu_{\rm Q}$ at the As site is more than 50 MHz.
     This would require a magnetic  field of more than $\sim$ 30 T (corresponding to the NMR frequency of $\sim$ 218 MHz) which is not easily accessible. 
      In addition, small changes in $\nu_{\rm Q}$ and ${\eta}$ produce a change in $K$, making the determination of $K$ difficult from the simulation of NMR spectrum.
% Furthermore, such a strong magnetic field would suppresses the ferromagntic spin fluctuations resulting in the difficulty in comparing with the zero-field electronic properties; for example, $T_1$ results in zero-field NQR. 
     Here, without using the usual method, we have succeeded in obtaining the Knight shift data from NQR spectrum under small magnetic fields lower than 0.5 T. 
     A similar measurement has been performed on the filled skutteruride compounds $A$Fe$_4$Sb$_{12}$ ($A$ = Sr, Ca) by Sakurai  ${\it et~al.}$ \cite{K}.  
  %    In such condition, the zeman interaction is more than one order smaller than the quadrupole interaction, thus it can be regarded as a perturbation to the quadrupole interaction.
     In the case of $\eta$ = 0, the NQR resonance frequency [$\nu_{\rm NQR}(H)$] under a small magnetic field can be written by \cite{Dean1954}
\begin{equation}
    \nu_{\rm NQR}(H) = \nu_{\rm NQR}(0) \pm \frac{\gamma_{\rm N}}{2\pi} (1+K)HF(\theta),
\label{eq:NQR_1}
\end{equation}
where  $\nu_{\rm NQR}(0)$ is  $\nu_{\rm NQR}$ at $H$ =  0 and $F(\theta)$ = $\frac{{\rm cos}\theta}{2}$$[3-(4{\rm tan}^2\theta+1)^{1/2}]$. 
    Under magnetic fields, the random distribution of $\theta$  produces the rectangular shape of the powder-pattern spectrum where $\theta$ = 0 (and also  $\pi$) produce both higher- and lower-frequency edges. 
    By measuring the external magnetic field dependence of the edge position of the NQR spectrum,  one can determine the coefficient of the second term of eq. (\ref{eq:NQR_1}), $\frac{\gamma_{\rm N}}{2\pi} (1+K)$, and thus the Knight shift since the $\frac{\gamma_{\rm N}}{2\pi}$ value is known.
     In the case of $\eta \neq$ 0, $\nu_{\rm NQR}(H)$ is given by 
\begin{equation}
    \nu_{\rm NQR}(H) = \nu_{\rm NQR}(0) \pm \frac{\gamma_{\rm N}}{2\pi} A(\eta) (1+K)HF(\theta).
\label{eq:NQR_2}
\end{equation}
    Here  one has a factor $A(\eta)$ in the second term, which depends on the value of $\eta$. 
    $A(\eta)$ is not equal to unity due to the mixing of the eigen states  $|$$m\rangle$ and  $|$$m\pm2\rangle$  originating from the finite value of $\eta$ in ${\cal H}_Q$. 
    In order to determine $K$, one needs to know the value of $A(\eta)$.  
    To determine  the $A(\eta)$ value, we have calculated the resonance frequency corresponding to the lower-frequency edge position $f_{\rm L}$ [see Fig. 3(a)]  under different magnetic fields by strictly diagonalizing ${\cal H}$ for $\eta$ = 0.4.  
   This calculation provides a value of $\frac{\gamma_{\rm N}}{2\pi} A(\eta)$ = 7.1417.
  Then we determined $A(\eta)$ to be 0.9794 using $\gamma_{\rm N}/2\pi$ = 7.2919 MHz/T for $^{75}$As nucleus.

\begin{figure}[tb]
\includegraphics[width=\columnwidth]{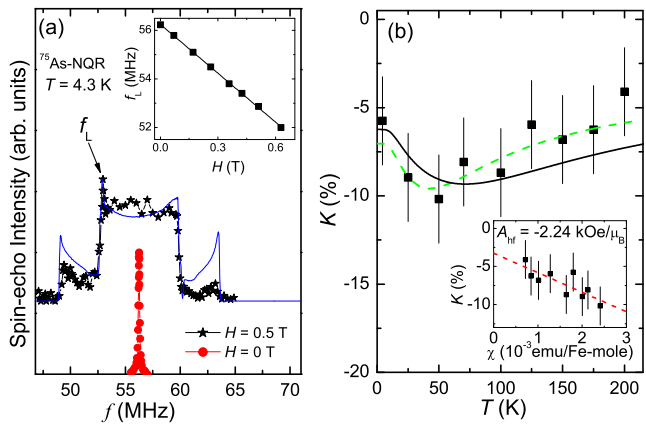} 
\caption{ (a) Representative $^{75}$As-NQR spectra under zero and 0.5 T magnetic field at 4.3 K.  The blue curve is a simulated powder-pattern spectrum with $\nu_{\rm Q}$ = 54.8 MHz, $\eta$ = 0.4 and $H$ = 0.5 T. 
   The  arrow shows the position of the lower-frequency edge position ($f_{\rm L}$) whose position can be calculated from  $\theta$ = $\pi$ (and also $\theta$ = 0). 
The inset shows the external magnetic field dependence of $f_{\rm L}$. 
 (b) $T$ dependence of  $^{75}$As Knight shift $K$.   The solid curves are the calculated results (see, text). 
 The inset shows $K$($T$) versus magnetic susceptibility $\chi$($T$).  The red dash line is a linear fit.
}
\label{fig:K}
\end{figure}

   Figure\ \ref{fig:K} (a) shows typical NQR spectra observed at $H$ =  0 and 0.5 T at  $T$ = 4.3 K where the rectangular shape of the powder pattern spectrum under $H$ = 0.5 T is clearly seen. 
    The small peaks (at $\sim$ 50 and 63 MHz) on both sides of the central rectangular spectrum are due to the mixing of states $|$$1/2\rangle$ and $|$$-1/2\rangle$ as a result of zero-order mixing effect \cite{PQR}.
    In fact, these features of the observed spectrum are relatively well reproduced by the calculated powder-pattern spectrum  with $\nu_{\rm Q}$ = 54.8 MHz, $\eta$ = 0.4 and $H$ = 0.5 T, as shown by the blue curve in Fig. \ \ref{fig:K} (a).
    The inset of Fig.\ \ref{fig:K} (a) shows a typical magnetic field dependence of $f_{\rm L}$  at the lower edge position (indicated by the black arrow), exhibiting a clear linear behavior.
    From the slope of -6.73 $\pm$ 0.18 MHz/T, the Knight shift $K$ at 4.3 K was determined to be -5.7 $\pm$  2.5 $\%$.  
   Although the error $\sim$ 40 $\%$ is relatively large, this is much better than the case of NMR spectrum from which we could not determine $K$. 
    It is also noted that we did not include any anisotropy in the Knight shift in the calculated spectrum which reproduces the observed one as shown above. 
   This suggests that, although one expects an anisotropic part in the Knight shift due to the local symmetry of the Fe ions (trigonal), 
 the anisotropy is not significant and could not be detected within our experimental uncertainty. 
    Therefore, the Knight shift discussed below is considered as an isotropic part of Knight shift.

  Following the above method,  we determined $K$ for each temperature  whose temperature dependence is shown in Fig.  \ \ref{fig:K} (b).
%    The negative value of $K$ originates from the negative hyperfine coupling constant.
 %   The experimental uncertainty is around 2$\%$, which is larger than the typical value estimated from conventional  NMR measurements.
 %   The uncertainty mainly comes from the linewidth of NQR spectra. 
    The temperature dependence of $K$ seems to be similar to that of $\chi$ which shows a broad maximum at $T^\ast$ $\sim$ 50 K. 
    This maximum is observed as a minimum in the $K$ data due to the negative hyperfine coupling constant as described below.
     The NMR shift consists of temperature dependent spin shift $K_{\rm s}(T)$ and $T$ independent orbital shift $K_{\rm 0}$; $K(T)$ = $K_{\rm s}(T)$ + $K_{\rm 0}$ where $K_{\rm s}(T)$ is proportional to the spin part of magnetic susceptibility  $\chi_{\rm s}$($T$) via hyperfine coupling constant $A_{\rm hf}$,  $K_{\rm s}(T)$  = $\frac{zA_{\rm hf}\chi_{\rm s}(T)}{N_{\rm A}}$.  
   Here  $N_{\rm A}$ is Avogadro's number and $z$ = 2 is the number of the nearest neighbor Fe ions at the As site.  
   The hyperfine coupling constant  is estimated to be $A_{\rm hf}$ =  -2.24 $\pm$ 0.6  kOe/$\mu_{\rm B}$  from the slope in the so-called $K$-$\chi$ plot shown in  the inset of  Fig.\ \ref{fig:K} (b).
   The obtained $A_{\rm hf}$ is comparable to -1.87 kOe/$\mu_{\rm B}$ at the Sb site reported in SrFe$_4$Sb$_{12}$ and CaFe$_4$Sb$_{12}$ \cite{K}.
   The transferred hyperfine coupling is due to the  coupling between Fe-3$d$ spins and As nucleus through the Fe-As covalent bond.
    $K_0$ is estimated to be -3.3 $\pm$ 1.1  $\%$.  
    Here we assumed that the observed $\chi(T)$  is almost ascribed to the spin part $\chi_{\rm s} (T)$ since the temperature independent part of the susceptibility $\chi_0$  is estimated to be less than 1 $\%$  of $\chi(T)$ even at 300 K from the analysis based on a Curie-Weiss fitting of the $\chi(T)$ data.

\begin{figure}[tb]
\includegraphics[width=\columnwidth]{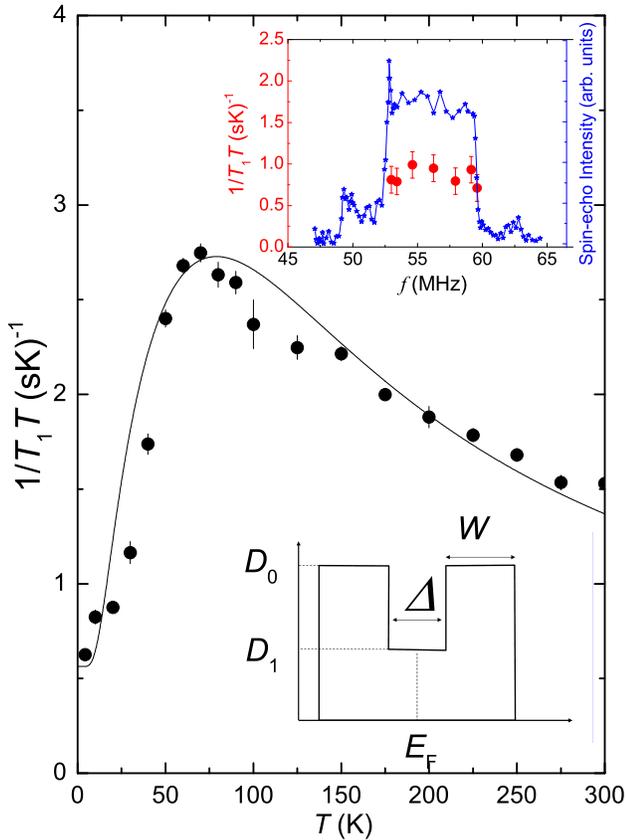} 
\caption{Temperature dependences of 1/$T_1T$ in SrFe$_4$As$_{12}$. 
  The solid line is the calculated result based on the a simple band structure shown in the lower inset with a set of parameters of  $\Delta$ = 88 K, $W$ = 220 K, and $r$ $\equiv$ $D_1$/$D_0$ = 0.38. 
  The upper inset shows the frequency dependence of 1/$T_1T$ measured at $T$ = 4.3 K and $H$ = 0.5 T, together with the spectrum at the same condition. No obvious change in the frequency dependence of 1/$T_1T$ indicates that 1/$T_1$ is nearly isotropic within the experimental uncertainty.}
\label{fig:T1}
\end{figure}

\subsection{C.  $^{75}$As spin-lattice relaxation rate 1/$T_1$}

        Figure \ \ref{fig:T1} shows the temperature dependence of the $^{75}$As  spin-lattice relaxation rate divided by temperature (1/$T_1T$) measured at the peak positions of the NQR spectra under zero magnetic field.
  %      1/$T_1T$ does not follow the Korringa law, $T_1T$ = constant, suggesting strong spin fluctuation in SrFe$_4$As$_{12}$.
        1/$T_1T$ increases gradually with decreasing temperature, and exhibits a broad maximum at $T$ $\sim$ 70 K, slightly higher than $T^\ast \sim$ 50 K observed in $\chi$.
         Below $T$ $\sim$ 70 K,  1/$T_1T$ decreases rapidly with decreasing temperature.
         Similar suppression of 1/$T_1T$ at low temperatures is observed in many filled skutterudite compounds, which has been discussed in terms of pseudo-gap  behavior at low temperatures \cite{Magishi2014,Matsumura2005,K,Matsumura2007,Toda2008}. %,Singh1994,Llunell1996,Fornari1999,Koga2005,Sichelschmidt2006}. 
       In order to analyze the temperature dependence of $1/T_1T$, we calculated  it based on a simple model where we adopt a concave-shaped band structure shown in the inset of Fig. \ \ref{fig:T1}.
      In this model, the Fermi energy ($E_{\rm F}$) is assumed to be at the center of the dip, and $\Delta$, $W$, the density of states ${\cal D}_0$ and ${\cal D}_1$ characterize the band structure near $E_{\rm F}$. 
    A similar calculation for CePt$_4$Ge$_{12}$  has been reported previously \cite{Toda2008}. 
      Using the formula, 
\begin{equation}
\frac{1}{T_1}\sim\int_0^\infty{\cal D}^2(E)f(E)(1-f(E)){\rm{d}}E 
\label{eq:T1}
\end{equation}
where $f(E)$ is the Fermi distribution function, we calculated $1/T_1T$ with a set of parameters of $\Delta$ = 88 K, $W$ = 220 K, and $r$ $\equiv$  ${\cal D}$$_1$/${\cal D}$$_0$ = 0.38, which reasonably reproduces the experimental data, as shown by the black curve in Fig. \ \ref{fig:T1}.  
   This indicates that the suppression of $1/T_1T$ below $T^\ast$ can be explained by the rigid  peculiar band structure modeled. 
% without introducing an opening of  pseudo-gap at low temperatures.
    We also calculated the temperature dependence of $K$ using the same model.  
    The solid line in Fig.  \ \ref{fig:K} (b) is the calculated result utilizing  the same set of parameters used for the $1/T_1T$ fitting, where $K_0$ = -3.3 $\%$ is used.
    Although the fit is not quite satisfactory, the model seems to capture the behavior of $K$ qualitatively.
     It is noted that a slightly  different set of parameters ($\Delta$ = 88 K, $W$ = 100 K, and $r$ = 0.38) improves the fitting as shown by the green broken line in Fig.  \ \ref{fig:K} (b).
    At present, it is not clear the reason why the different sets of parameters are needed to fit each data set well.
    Since our model is very crude, one could improve the fit if more detailed band structure were  available.  
    Further studies, such as band calculation and angle-resolved photoemission spectroscopy measurement will be needed to confirm the peculiar band structure proposed from the present study and also obtain the detailed information of  the band structure.

    \subsection{D.  Ferromagnetic spin fluctuations}

\begin{figure}[tb]
\includegraphics[width=\columnwidth]{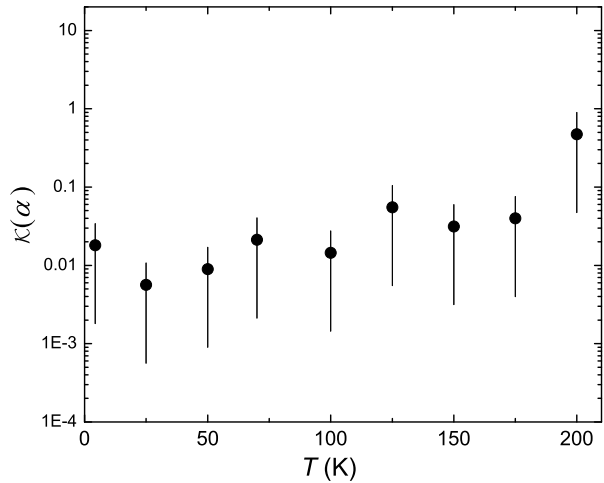} 
\caption{ $T$ dependence of Korringa ratio ${\cal K}$($\alpha$)  in SrFe$_4$As$_{12}$.
% The inset shows the temperature dependence of 1/$T_1TK_{\rm s}$ (filled red squares)  and 1/$T_1TK_{\rm s}^{3/2}$ (filled blue circles). 
}
\label{fig:Alpha}
\end{figure}

      Now we discuss the magnetic fluctuations in SrFe$_4$As$_{12}$ based on the $T_1$ and the spin part of the Knight shift ($K_{\rm s}$) data.
     Here we apply a Korringa ratio analysis to extract the character of spin fluctuations.
     Within a Fermi liquid picture, $1/T_1T$ is proportional to the square of the density of states  at the Fermi energy ${\cal D}$($E_{\rm F}$) and $K_{\rm s}$  is proportional to ${\cal D}$($E_{\rm F}$), leading to the Korringa relation of $T_1TK_{\rm s}^2$  = $\frac{\hbar}{4\pi k_{\rm B}} \left(\frac{\gamma_{\rm e}}{\gamma_{\rm N}}\right)^2$ = ${\cal S}$. 
     Here  $\gamma_{\rm e}$ is the electron gyromagnetic ratio. 
    For the $^{75}$As nucleus,  ${\cal S} =8.97\times 10^{-6}$ Ks. 
    The Korringa ratio ${\cal K}$$(\alpha)\equiv$ $\frac{2 \cal S}{T_1TK_{\rm s}^2}$, which reflects the deviations from ${\cal S}$, can reveal information about electron correlations in the material \cite{Moriya1963,Narath1968}.
     It is noted that we have included a factor of 2 in the Korringa ratio by taking the effect of the number of the nearest neighbor Fe irons into consideration. 
    ${\cal K}$$(\alpha)$ $\sim1$ represents the situation of uncorrelated electrons.  
     However, enhancement of  $\chi$ ($\mathbf{q}\neq$ 0) increases 1/$T_1T$ but has little or no effect on $K_{\rm s}$, which probes only the uniform $\chi$ ($\mathbf{q}$ = 0). 
       Thus ${\cal K}$$(\alpha)$ $>$ 1 indicates antiferromagnetic (AFM)  spin correlations. 
      In contrast, FM spin correlations produces ${\cal K}$$(\alpha)$ $<$ 1.
%     These come from the enhancement of $\chi(\mathbf{q}\neq 0)$, which increases $1/T_1T$ but has little or no effect on $K_{\text{spin}}$, since the latter probes only the uniform $\chi(\mathbf{q} = 0)$.  
   Therefore, the predominant feature of magnetic correlations, whether AFM or FM, can be determined by the Korringa ratio ${\cal K}$$(\alpha)$. 
    Figure \ \ref{fig:Alpha} shows the temperature dependence of the Korringa ratio ${\cal K}$$(\alpha)$.
   ${\cal K}$$(\alpha)$ seems to be nearly temperature independent with most values around 0.02, although the value of ${\cal K}$$(\alpha)$ at 200 K is slightly higher than others. 
   ${\cal K}$$(\alpha)$ values much smaller than unity clearly evidence dominant ferromagnetic fluctuations in SrFe$_4$As$_{12}$.
    A similar, but slightly smaller, value of ${\cal K}$$(\alpha)$ = 0.005 has been reported in SrFe$_4$Sb$_{12}$ and CaFe$_4$Sb$_{12}$ \cite{K}.
    These results indicate that the change in ligand does not significantly affect the magnetic properties of the skutterudite compounds, although the magnetic properties largely depend on the number of the valence electrons of the $A$ ions in $A$Fe$_4X_{12}$ as described in Introduction.
  
   In this analysis, we have assumed that 1/$T_1$ is isotropic. 
     In order to check the anisotropy in 1/$T_1$, we have measured the position dependence of 1/$T_1$ in the As NQR spectrum under a magnetic field of 0.5 T. 
      As shown in the upper inset of Fig. 4, we do not observe significant change in 1/$T_1$  at different positions (i.e., different $\theta$)  of the spectrum within our experimental uncertainty, indicating that $T_1$ is nearly isotropic. 
     In addition, it should be noted that the observed 1/$T_1T$ is the sum of two contributions: the spin and orbital relaxation rates. 
     As a result, the estimated values for ${\cal K}$$(\alpha)$  can be considered to be upper limits  on ${\cal K}$$(\alpha)$, indicating even stronger FM correlations in SrFe$_4$As$_{12}$ than expected from the above ${\cal K}$$(\alpha)$ values.

 %    According to  self-consistent renormalization (SCR) theory,  1/$T_1T$ is proportional to $K_{\rm s}$ or to $K^{3/2}_{\rm s}$ for three dimensional (3D) or  two-dimensional (2D) ferrromagnetic spin fluctuations, respectively, for  weak itinerant ferromagnets \cite{SCR1, SCR2}.   The inset of Fig.~\ref{fig:Alpha} shows the temperature dependence of 1/$T_1TK_{\rm s}$ and 1/$T_1TK_{\rm s}^{3/2}$.      We could not find any clear difference between the plots within our experimental uncertainty which mainly comes from the error bar in the $K$ data.     Therefore, we could not determine the dimensionality of the FM spin fluctuation from the experimental data.  Further studies under high magnetic fields for more precise determination of $K$ may help to characterize the dimensionality of the spin fluctuations of this system, although high magnetic fields may affect the nature of spin fluctuations.   

   \section{IV. Summary}
      In summary,  we have carried out $^{75}$As NMR and NQR measurements on the filled skutterudite SrFe$_4$As$_{12}$.
      No broadening or splitting of the NQR spectra observed around 56 MHz confirms no phase transition in SrFe$_4$As$_{12}$. 
     Using the NQR spectra under small magnetic field, we have succeeded in determining  Knight shift $K$ whose temperature dependence shows a  minimum at $T^\ast \sim$ 50 K corresponding to the maximum in the magnetic susceptibility $\chi$.
    The nuclear spin-lattice relaxation rate divided by temperature, 1/$T_1T$, also exhibits a broad maximum at $T$ $\sim$ 70 K, slightly higher than $T^\ast$. 
    Those temperature dependences have been reasonably explained by the simple model in which a concave-shaped band structure near the Fermi energy is assumed.
   The  Korringa ratio parameter ${\cal K}$$(\alpha)$ is found to be much smaller than unity, revealing  the existence of  ferromagnetic fluctuations.
    Our NMR data clearly indicate  that SrFe$_4$As$_{12}$  is a metal with ferromagnetic spin correlations. 
    Similar concave-shaped DOS has been observed in other skutterudite compounds which do not exhibit ferromagnetic order although strong ferromagnetic fluctuations have been revealed \cite{Matsumura2007,Magishi2014}. These results may suggest that the characteristic concave-shaped DOS would be relevant to the absence of ferromagnetic order in those systems.
    Finally, it is worth to mention  that the isostructural compound SrOs$_4$As$_{12}$ is a superconductor with a $T_{\rm c}$ of 4.8 K \cite{Nishine2017}. 
    Quite recently a similar compound with the same structure, CaOs$_4$P$_{12}$, has been suggested to be a BCS-type superconductor with $T_{\rm c}$ = 2.5 K \cite{Kawamura2018}.
        It would be very interesting to perform NMR and NQR measurements on SrOs$_4$As$_{12}$ and CaOs$_4$P$_{12}$ for investigating  their magnetic  fluctuations, as well as superconducting properties, from a microscopic point of view.

   \section{Acknowledgments} 
    The authors thank Mamoru Yogi for helpful discussions.  
The research was supported by the U.S. Department of Energy (DOE), Office of Basic Energy Sciences, Division of Materials Sciences and Engineering. Ames Laboratory is operated for the U.S. DOE by Iowa State University under Contract No.~DE-AC02-07CH11358.
    Part of this work was supported by JSPS KAKENHI Grant Number 23340092.


\begin{thebibliography}{10}
\bibitem{Jeitschko1977} W. Jeitschko and D. Braun, Acta Crystallogr., Sect. B {\bf 33}, 3401 (1977).
\bibitem{Sales2005} B. C. Sales, Filled skutterudites, in $Handbook$ $on$ $the$ $Physics$ $and$ $Chemistry$ $of$ $the$ $Rare$ $Earths$, edited by K. A. Gschneidner Jr., J.-C. Bunzli, and V. K. Pecharsky (Elsevier Science, Amsterdam, 2003), Vol. 33, Chap. 211, and references there in.
\bibitem{Shirotani1997} I. Shirotani, T. Uchiumi, K. Ohno, C. Sekine, Y. Nakazawa, K. Kanoda, S. Todo, and T. Yagi, Phys. Rev. B {\bf 56}, 7866 (1997).
\bibitem{Sekine1998} C. Sekine, T. Uchiumi, I. Shirotani, and T. Yagi, Phys. Rev. Lett. {\bf 79}, 3218 (1997).
\bibitem{Leithe-Jasper2003} A. Leithe-Jasper, W. Schnelle, H. Rosner, N. Senthilkumaran, A. Rabis, M. Baenitz, A. Gippius, E. Morozova, J. A. Mydosh, and Y. Grin, Phys. Rev. Lett. {\bf 91}, 037208 (2003).
\bibitem{Schnelle2008} W. Schnelle, A. Leithe-Jasper, H. Rosner, R. Cardoso-Gil, R. Gumeniuk, D. Trots, J. A. Mydosh, and Y. Grin, Phys. Rev. B {\bf 77}, 094421 (2008).
\bibitem{Ishida2005} K. Ishida, H. Murakawa, K. Kitagawa, Y. Ihara, H. Kotegawa, M. Yogi, Y. Kitaoka, Ben-Li Young, M. S. Rose, D. E. MacLaughlin, H. Sugawara, T. D. Matsuda, Y. Aoki, H. Sato, and H. Harima, Phys. Rev. B {\bf 71}, 024424 (2005).
\bibitem{Matsuoka2005} E. Matsuoka, K. Hayashi, A. Ikeda, K. Tanaka, T. Takabatake, and M. Matsumura, J. Phys. Soc. Jpn. {\bf 74}, 1382 (2005).
\bibitem{Matsumura2005} M. Matsumura, G. Hyoudou, H. Kato, R. Nishioka, E. Matsuoka, H. Tou, T. Takabatake, and M. Sera, J. Phys. Soc. Jpn. {\bf 74}, 2205 (2005).
\bibitem{Nakai2005} Y. Nakai, K. Ishida, D. Kikuchi, H. Sugawara, and H. Sato, J. Phys. Soc. Jpn. {\bf 74}, 3370 (2005)
\bibitem{Matsumura20071} M. Matsumura, H. Kato, T. Nishioka, E. Matsuoka, K. Hayashi, and T. Takabatake, J. Magn. Magn. Mater. {\bf 310}, 1035 (2007).
\bibitem{Shimizu2007} M. Shimizu, H. Amanuma, K. Hachitani, H. Fukazawa, Y. Kohori, T. Namiki, C. Sekine, and I. Shirotani, J. Phys. Soc. Jpn. {\bf 76}, 104705 (2007).
\bibitem{Tou2011} H. Tou, Y. Inaoka, M. Doi, M. Sera, K. Asaki, H. Kotegawa, H. Sugawara, and H. Sato, J. Phys. Soc. Jpn. {\bf 80}, 074703 (2011).
\bibitem{Magishi2014} K. Magishi, R. Watanabe, A. Hisada, T. Saito, K. Koyama, J. Phys. Soc. Jpn. {\bf 83}, 84712 (2014).
\bibitem{Kawamura2018} Y. Kawamura, S. Deminami, L. Salamakha, A. Sidorenko, P. Heinrich, H. Michor, E. Bauer, and C. Sekine, Phys. Rev. B {\bf 98}, 024513 (2018).  
\bibitem{Sales1996} B. C. Sales, D. Mandrus, and R. K. Williams, Science {\bf 272}, 1325 (1996).


\bibitem{Leithe-Jasper2004} A. Leithe-Jasper, W. Schnelle, H. Rosner, M. Baenitz, A. Rabis, A. A. Gippius, E. N. Morozova, H. Borrmann, U. Burkhardt, R. Ramlau, U. Schwarz, J. A. Mydosh, Y. Grin, V. Ksenofontov, and S. Reiman, Phys. Rev. B {\bf 70}, 214418 (2004).

\bibitem{Schnelle2005} W. Schnelle, A. Leithe-Jasper, M. Schmidt, H. Rosner, H. Borrmann, U. Burkhardt, J. A. Mydosh, and Y. Grin, Phys. Rev. B {\bf  72}, 020402(R) (2005).
\bibitem{K} A. Sakurai, M. Matsumura, H. Kato, T. Nishioka, E. Matsuoka, K. Hayashi, and T. Takabatake, J. Phys. Soc. Jpn. {\bf 77}, 063701 (2008). 
\bibitem{Gippius2006} A. Gippius, M. Baenitz, E. Morozova, A. Leithe-Jasper, W. Schnelle, A. Shevelkov, E. Alkaev, A. Rabis, J. Mydosh, Y. Grin, and F. Steglich, J. Magn. Magn. Mater. {\bf 300}, 403(E) (2006).

\bibitem{Nishine2017} K. Nishine, Y. Kawamura, J. Hayashi, and C. Sekine, Jpn. J. Appl. Phys. {\bf 56}, 05FB01 (2017).


%\bibitem{Matsuoka2006} E. Matsuoka, S. Narazu, K. Hayashi, K. Umeo, and T. Takabatake, J. Phys.Soc. Jpn. {\bf75}, 014602 (2006).
%\bibitem{T1}  $1/T_1$ at each $T$ was determined by fitting the nuclear magnetization $M$ versus time $t$  using the exponential function $1-M(t)/M(\infty) = e^ {-3t/T_{1}}$ for $^{75}$As NQR,  where $M(t)$ and $M(\infty)$ are the nuclear magnetization at time $t$ after the saturation and the equilibrium nuclear magnetization at $t$ $\rightarrow$ $\infty$, respectively. 
\bibitem{Sekine2015} C. Sekine, T. Ishizaka, K. Nishine, Y. Kawamura, J. Hayashi, K. Takeda, H. Gotou, and Z. Hiroi, Phys. Procedia {\bf 75}, 383 (2015).
\bibitem{impurity} $^{75}$As NQR measurements of the impurity phase show that NQR frequency is $\nu_{\rm Q}$ = 23.6 MHz, $\eta$ = 0, and 1/$T_1T \sim$ 0.5 (sK)$^{-1}$ from 4.3 K up to 150 K.  Since these values are close to $\nu_{\rm Q}$ = 22.757 MHz and  1/$T_1T \sim$ 0.68 (sK)$^{-1}$ in arsenic metal \cite{AsMetal1, AsMetal2}, the impurity phase could be assigned to arsenic metal.

\bibitem{AsMetal1} T. J.  Bastow, J. Phys.: Condens. Matter {\bf 11}, 569 (1999).
\bibitem{AsMetal2} D. E. Jellison and P. C. Taylor,  Solid State Commun. {\bf 27}, 1025 (1978).

\bibitem{Tou2005} H. Tou, M. Doi, M. Sera, M. Yogi, Y. Kitaoka, H. Kotegawa, G.-q. Zheng, H. Harima, H. Sugawara, H. Sato, Physica B {\bf 359-361}, 892 (2005).
\bibitem{Yogi2009} M. Yogi, H. Niki, M. Yashima, H. Mukuda, Y. Kitaoka, H. Sugawara, and H. Sato, J. Phys. Soc. Jpn. {\bf 78}, 053703 (2009).
\bibitem{Matsumura2007} M. Matsumura, G. Hyoudou, M. Itoh, H. Kato, T. Nishioka, E. Matsuoka, H. Tou, T. Takabatake, and M. Sera, J. Phys. Soc. Jpn. {\bf 76}, 084716 (2007).
\bibitem{Nowak2009}  B. Nowak, O. {\.Z}oga\l, A. Pietraszko, R. E. Baumbach, M. B. Maple, and Z. Henkie, Phys. Rev. B {\bf 79}, 214411 (2009).
\bibitem{Nowak2011} B. Nowak, O. {\.Z}oga\l, Z. Henkie, and M.B. Maple, Solid State Communications  {\bf 151}, 550 (2011).
\bibitem{Yogi2014} M. Yogi, H. Niki, T. Kawata, and C. Sekine, JPS Conf. Proc. {\bf 3},  011046 (2014).
\bibitem{alphaQ} S. Takagi, H. Muraoka, T. D. Matsuda, Y. Haga, S. Kambe, R. E. Walstedt, E. Yamamoto, and Y. $\bar {\rm O}$nuki, J. Phys. Soc. Jpn. {\bf 73},  469 (2004).
\bibitem{Dean1954}C. Dean, Phys. Rev. {\bf 96}, 1053 (1954).

\bibitem{PQR} T. P. Das and E. L. Hahn, Nuclear Quadrupole Resonance Spectroscopy in Solid State Physics, Supplement I (Academic Press Inc., New York, 1958).
\bibitem{Toda2008} M. Toda, H. Sugawara, K. Magishi, T. Saito, K. Koyama, Y. Aoki, H. Sato, J. Phys. Soc. Jpn. {\bf 77}, 124702  (2008).
\bibitem{Moriya1963} T. Moriya, J. Phys. Soc. Jpn. {\bf 18}, 516 (1963).
\bibitem{Narath1968} A. Narath and H. T. Weaver, Phys. Rev. {\bf 175}, 373 (1968).
%\bibitem{Moriya1995} M. Hatatani and T. Moriya, J. Phys. Soc. Jpn. {\bf 64}, 3434 (1995).
%\bibitem{FeSi} M. Corti, S. Aldovandi, M. Funciulli, and F. Tabak: Phys. Rev. B {\bf 67}, 172408 (2003).  
%\bibitem{SCR1} M. Hatatani and T. Moriya, J. Phys. Soc. Jpn. {\bf 64}, 3434 (1995).
%\bibitem{SCR2} T. Moriya, Spin~Fluctuations~in~Itinerant~Electron~Magnetism, Vol. 56 (Springer-Verlag, 1985).

\end{thebibliography}
\end{document}